\documentclass{INTERSPEECH2023}

\interspeechcameraready

\title{a Mask Free Neural Network for Monaural Speech Enhancement}
\name{Liang LIU$^1$, Haixin GUAN$^1$$^2$, Jinlong MA$^1$, Wei DAI$^1$, Guangyong WANG$^1$, Shaowei DING$^1$}
\address{
 $^1$Unisound AI Technology Co. Ltd, China\\
 $^2$University of Science and Technology of China, China}
\email{liuliang, guanhaixin, majinlong, daiwei, wangguangyong, dingshaowei@unisound.com}

\begin{document}

\maketitle
 
\begin{abstract}
In speech enhancement, the lack of clear structural characteristics in the target speech phase requires the use of conservative and cumbersome network frameworks. It seems difficult to achieve competitive performance using direct methods and simple network architectures. However, we propose the MFNet, a direct and simple network that can not only map speech but also map reverse noise. This network is constructed by stacking global local former blocks (GLFBs), which combine the advantages of Mobileblock for global processing and Metaformer architecture for local interaction. Our experimental results demonstrate that our network using mapping method outperforms masking methods, and direct mapping of reverse noise is the optimal solution in strong noise environments. In a horizontal comparison on the 2020 Deep Noise Suppression (DNS) challenge test set without reverberation, to the best of our knowledge, MFNet is the current state-of-the-art (SOTA) mapping model.

\end{abstract}

\section{Introduction}

With the development of deep learning, Speech enhancement (SE) techniques have achieved significant progress. Typically, those can be divided into two categories, time domain methods \cite{stoller2018wavunet,luo2019conv} and T-F domain methods \cite{williamson2015crm,hu2020dccrn}. Especially, the latter one have obtained better performance in DNS Challenge \cite{reddy2020interspeech,reddy2021icassp,reddy2021interspeech,dubey2022icassp}, one of the most influential competitions in the field of SE. Therefore, the goal of this study is to design an effective T-F domain system for single channel speech enhancement.

In T-F domain speech enhancement methods, the direct learning of the T-F spectrum values (mapping method \cite{xu2014regression,tan2019GCRN}) and the learning of T-F masking (masking method \cite{hu2020dccrn,wang2014training}) are two classic methods. 
  Mapping the magnitude \cite{xu2014regression} or mapping the real and imaginary parts is a direct and radical approach, but it seems to be a difficult problem, so GCRN \cite{tan2019GCRN} requires two decoders to map the real and imaginary parts separately. The masking method simplifies the problem by starting from the prior of the noisy speech components. It is estimated either in the rectangular coordinate system DPCRN \cite{le2021dpcrn} FullSubnet \cite{hao2021fullsubnet} or in the polar coordinate system DCCRN \cite{hu2020dccrn} DCUNet \cite{DCUNET}. On this basis, DeepFilterNet \cite{schroter2022deepfilternet,schroter2022deepfilternet2} using nearby filtering and summation can slightly compensate for the theoretical defects of the masking method. 

As development progresses, the work of combining the two (referred to as  decoupling methods \cite{yin2020phasen,li2022taylor,ctsnet,HGCN+}) seems to be increasingly popular. For example, PHASEN \cite{yin2020phasen} decouples the task into magnitude masking and phase mapping, TaylorSENet \cite{li2022taylor} further generalize the decoupling method into two parts: magnitude estimation and complex estimation. CTSNet \cite{ctsnet} attempts to decouple the mapping method, that is, first mapping the magnitude spectrum and then mapping the complex spectrum. Moreover, researchers have pushed the limits of complexity by incorporating multiple stages of the decoupling approach, each leveraging a large cascading network (referred to as cascading network \cite{li2021iDMP,li2021sdd,zhao2022frcrn,wang2021tripledomain}). As a result, the total computation and number of parameters in the network grow exponentially with each additional stage. Although this approach can lead to improved performance and enable the network to learn more intricate features, it is crucial to consider the trade-off between performance and computational cost.


\begin{figure*}[!ht] 
\centering 
\includegraphics[width=1\textwidth]{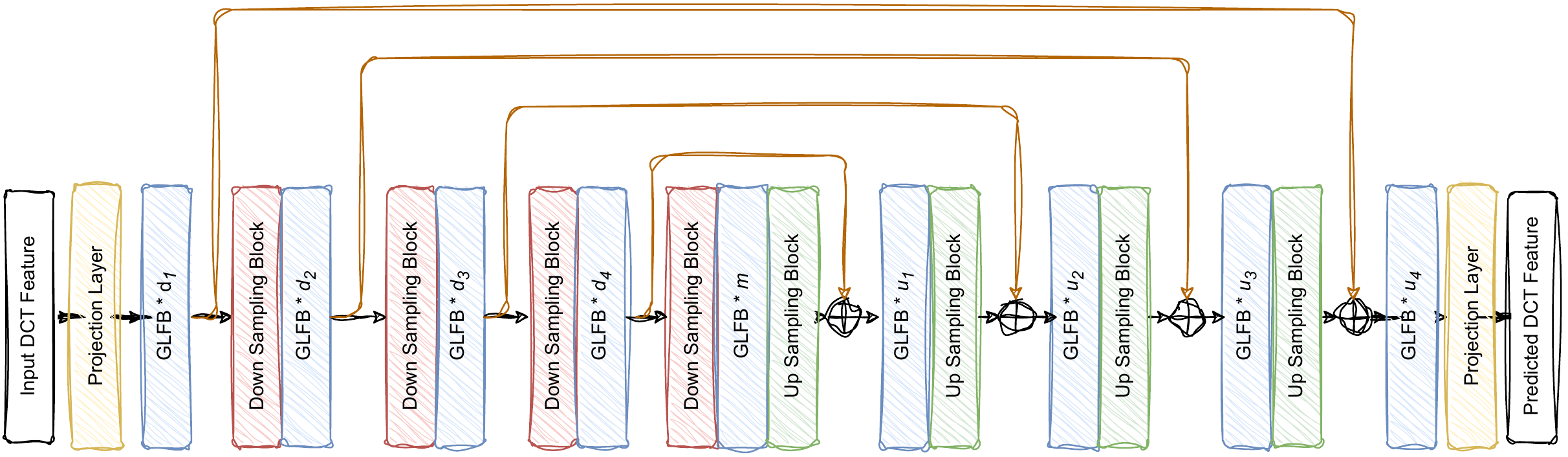} 
\caption{Architecture of the proposed MFNet} 
\label{Fig1} 
\end{figure*}


Through the above observations, we have identified some uncertainties and put forward hypotheses:

\begin{itemize}
    \item As the current trend in single-channel SE, it appears challenging to attain competitive performance using straightforward techniques and basic network architectures.
    \item There is a contradiction between the research results and the past studies \cite{DCUNET,wu2023rethinking} on which method performs better between masking and mapping. It seems that with the reasonable optimization of the network, the mapping method appears to be more direct and less aggressive.
    \item The decoupling method adopts a multi-step estimation strategy to solve the problem of phase estimation, which makes the overall process more complex. If the problem of phase estimation can be solved, the structure of the network will be greatly simplified.    
\end{itemize}

According to review \cite{overview}, all current training objectives can be collectively referred to as SA method. For example, the masking method can be expressed as  
$ L_{SA-masking} = || S-\hat{M}\cdot Y  || $, the mapping method can be expressed as
$ L_{SA-mapping} = ||  S_r-\hat{S}_r || + ||  S_i-\hat{S}_i ||  $, and the decoupling method can be expressed as
$ L_{SA-decoupling} = || \left | S \right | -\hat{M}\cdot \left | Y \right | || + || S_{r,i}-\hat{S}_{r,i} ||  $. 
Cascading method can be expressed as
$ L_{SA-cascading} = L_{stage1} + L_{stage2}  $.
In the above equation, $L$ represents the loss function, $S$ represents the target speech signal, $\hat{S}$ represents the predicted speech signal, $\hat{M}$ represents the predicted mask,  $Y$ represents the noisy speech signal, The subscripts $r$ and $i$ represent the real and imaginary parts respectively. Both $stage1$ and $stage2$ can be represented using either masking, mapping, or cascading. By observation, we believe that the above expressions can be unified into a intuitive way, 
$ L_{SA-intuitive} = || S - \hat{S}  || $. Based on this premise, we propose a simple single-stage neural network for speech enhancement that utilizes short-time discrete cosine transform (STDCT) \cite{rao2014discrete} features and does not require a mask. This network has the following characteristics:
\begin{itemize}
    \item We have designed an efficient and lightweight module called GLFB, which is based on the structural features of MetaFormer architecture \cite{yu2022metaformer}, MobileNet block \cite{sandler2018mobilenetv2}, and design experience from NAFNet \cite{chen2022simple}. The module prototype is based on MetaFormer, with global modeling accomplished using depth-wise separable convolution, gating mechanism, and channel attention mechanism. Local modeling is done by point convolution.

\item Our network structure is simple, consisting of three modules: encoder, decoder, and bottleneck, each of which is composed of GLFB. The encoder utilizes small-sized convolution kernels for down-sampling, while the decoder employs pixel-shuffle method for up-sampling. We establish jump layer connections by direct summation.

\item Our proposed network uses real-valued STDCT spectrum as its input features. Unlike STFT features, which require complex values, STDCT features are represented solely using real values, resulting in a more uniform representation. The network is designed to perform speech enhancement without learning a mask, making it capable of mapping both speech and reverse noise. We named the network MFNet.
\end{itemize}

Our experimental results demonstrate that our proposed network outperforms the masking approach when using the mapping approach. We also discovered an interesting result where our network achieves better performance in a strong noise environment when directly learning the reverse noise compared to mapping the speech. On the DNS 2020 test set without reverberation, our proposed model achieves a fairly competitive performance. Based on our current understanding, in mapping method, this model performs the best on the given test set. 

The rest of this paper is organized as follows. Section 2 introduces the proposed method. Section 3 describes the experiments and results. Section 4 is a comprehensive conclusion.

\section{Proposed method}

\subsection{STDCT input feature}
We have utilized the STDCT spectrum as our input feature, which is a real-valued transformation that preserves all the information present in the signal and contains implicit phase information\cite{gao2021low,DCTCRN}. This eliminates the need for designing a complex neural network to estimate the explicit phase of the signal, which can be challenging and computationally expensive. Additionally, using the STDCT spectrum removes the necessity of estimating the complex mask that is required in some other audio processing techniques. 

\subsection{Model architecture}
We adopted a UNet-shaped network architecture because it is suitable for intensive prediction tasks at the T-F bins level. The network structure of our model is designed to be as parsimonious as possible and the modules are designed to be reusable, taking inspiration from the design concept of NAFNet. The model has three parts: encoder, bottleneck layer, and decoder. From the perspective of the included base modules, the network as a whole contains only three modules: projection layer, GLFB, and sampling (down or up). The encoder contains the projection layer, GLFB, and down-sampling modules. The bottleneck layer contains only GLFB. The decoder contains the projection layer, GLFB, and up-sampling modules. The projection layer on the input side projects the STDCT features into high-dimensional space, keeping the size of the feature map unchanged but increasing the number of channels from one to the number of channels set by the model $n$. The number of channels in the feature map is doubled for each down-sampling layer, and halved for each up-sampling layer, resulting in an overall number of channels for the model of $[n, 2n, 4n, 8n, 16n, 8n, 4n, 2n, n]$. Notably, there is no activation function used in the entire network. The performance of the network is mainly determined by the stacking of GLFB. The features extracted from the encoder stage are added directly to the decoder stage instead of the usual concatenate practice, which reduces the number of parameters in the decoder stage by reducing the number of convolution kernel groups. The encoder, bottleneck, and decoder each have a number of blocks represented by $[d1, d2, d3, d4], [m], [u1, u2, u3, u4]$. The overall network structure is shown in Figure \ref{Fig1}, and the text length remains similar.

\begin{figure}[ht] 
\centering 
\includegraphics[width=0.44\textwidth]{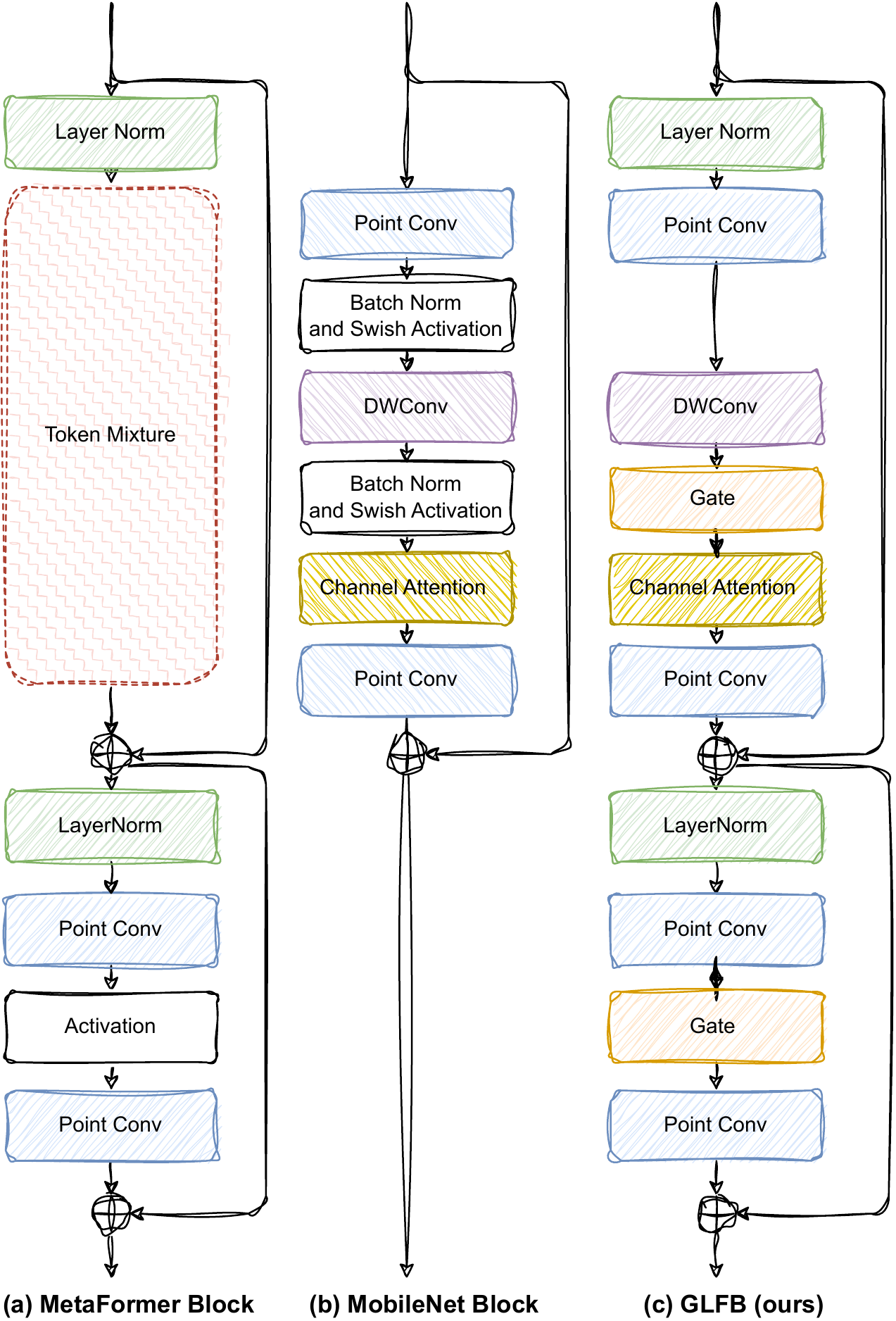} 
\caption{Architecture of the GLFB} 
\label{Fig2} 
\end{figure}

\subsection{Down-sampling, up-sampling and projection layer}
In UNet-shaped networks, down-sampling and up-sampling are typically accomplished through convolution and transpose convolution, respectively. However, some researchers have utilized larger convolution kernels to improve model performance, resulting in larger model parameters and increased computational effort. In contrast, the MFNet approach employs a convolution kernel size of 2 and a stride of 2 for down-sampling, while up-sampling is achieved using the pixel-shuffle operation to avoid the checkerboard grid effect that can occur with transposed convolution. The projection layer in MFNet uses a $3 \times 3$ convolution and is responsible for extracting features from a single-channel input, projecting them into high-dimensional features via input projection, and then projecting the features back into a single-channel output via output projection.

\subsection{Global local former block}
The GLFB is a crucial module in MFNet. It draws design inspiration from the transformer architecture \cite{vaswani2017attention}, which includes a multi-head attention module and a feed-forward network module. However, the vanilla self-attention mechanism suffers from quadratic computational complexity with the size of the feature map, making it unsuitable for mobile or resource-constrained devices. To tackle this problem, we adopt a modified MobileNet block as a replacement for the multi-head attention module, which is inspired by the research on Metaformer blocks. This approach solves the problem of complexity with token length dependence on $O(n^2)$. At the same time, this module has similar global and local modeling capabilities as transformer.  The global modeling part is done by depth-wise separable convolution, simple gating mechanism and the channel attention mechanism, and the local modeling part is done by point convolution. The feed-forward network module is slightly modified by replacing the activate layer to gate layer. The details are shown in Figure \ref{Fig2}.

The simple channel attention module used in our model is the same as the one used in the MobileNet Block. DWConv refers to depth-wise separable convolution, and Point Conv refers to point-wise convolution. In \ref{Fig2}(c), the module includes four Point Convs. The first and third Point Conv double the number of input channels, while the second and fourth Point Conv maintain the same number of channels. The gate mechanism halves the number of channels.

\subsection{Loss function}
We propose a loss function for MFNet. This loss function contains two components, the first one is the mean-square error (MSE) loss for absolute values of STDCT. This part is written as
\begin{align}
Loss_{abs} =  
||  \ | S_{STDCT}| -| \hat{S}_{STDCT}|   \ ||  _{2}^{2}, 
\end{align}
and the second one is the MSE loss for polar values. This part is written as
\begin{align}
Loss_{polar} =  
|| S_{STDCT} - \hat{S}_{STDCT}  ||  _{2}^{2} .
\end{align}
A hyper-parameter $\gamma $ is the weight to adjust the weight of absolute MSE and polar MSE contribution. The loss function of MFNet is written as
\begin{align}
Loss_{MFNet} = \gamma \cdot  Loss_{abs} + (1-\gamma )\cdot Loss_{polar}.
\end{align}

In the formula, $S$ represents the target speech signal, $\hat{S}$ represents the speech signal predicted by the network.

\begin{table}[]
\caption{Ablation study of mask-free method}
\centering
\label{tab2}
\begin{tabular}{@{}lccc@{}}
\toprule
\textbf{Model}               & \textbf{PESQ} & \textbf{STOI}  & \textbf{SNR}   \\ \midrule
DCTCRN \cite{DCTCRN}                & 2.80         & 0.863          & 11.55          \\
Cascade DCTCRN                 & 2.83          & 0.867          & 11.59          \\
TaylorSENet \cite{li2022taylor}                  & 2.92          & 0.877          & 11.79          \\
Ours(Masking)                   & 3.02 & 0.902          & 13.62          \\
Ours(Mapping Speech)         & 3.02          & 0.902          & 13.72          \\
Ours(Mapping Reverse Noise) & \textbf{3.05}          & \textbf{0.904} & \textbf{13.93} \\ \bottomrule
\end{tabular}
\end{table}

\begin{table*}[ht]
\caption{Experimental results on the DNS 2020 test set w/o reverberation}
\label{DNS table}
\centering
\begin{tabular}{@{}lcccccc@{}}
\toprule
\textbf{Model} &\textbf{Method}&  \textbf{MACs(G/s)} & \textbf{WB-PESQ} & \textbf{NB-PESQ} & \textbf{STOI} & \textbf{SI-SDR(dB)} \\ \midrule
\textbf{Noisy}     &              &       & 1.58 & 2.45  & 91.52 & 9.07  \\
\textbf{DCCRN(2020) \cite{hu2020dccrn}}  &masking    & 11.13 & -    & 3.27  & -     & -     \\
\textbf{FullSubNet(2021) \cite{hao2021fullsubnet}} & masking   & 31.35 & 2.78 & 3.31  & 96.11 & 17.29 \\
\textbf{CTSNet(2021) \cite{ctsnet}} & decoupled-based mapping   & 5.57    & 2.94 & 3.42 & 96.21 & 16.69 \\
\textbf{TaylorSENet(2022) \cite{li2022taylor}} & decoupled-based masking     & 6.14  & 3.22 & 3.59  & 97.36 & 19.15 \\
\textbf{FRCRN(2022) \cite{zhao2022frcrn}}   & cascading     & 241.98\footnote{We use the code provided by the website for statistical calculations}     & 3.23 & 3.60   & 97.69 & 19.78 \\
\midrule
\textbf{MFNet}   &mapping     & 6.09  & \textbf{3.43} &\textbf{ 3.74 } & \textbf{97.98} & \textbf{20.31}  \\ \bottomrule
\end{tabular}
\end{table*}


\section{Experiments and results}
\subsection{Datasets}

In the experiment, we used data from two datasets.

\textbf{DNS-Challenge}. The Interspeech 2020 DNS-Challenge corpus \cite{reddy2020interspeech} covers over 500 hours of clean clips by 2150 speakers and over 180 hours of noise clips. For model evaluation, it provides a non-blind validation set with two categories, namely with and without reverberation, and each includes 150 noisy-clean pairs. Following the scripts provided by the organizer, We generate 3000h data for training and the SNRs randomly range from -3dB to 15dB. To ensure fairness in the experiment, we used official scripts to generate data and did not use any data augmentation techniques. 

\textbf{TIMIT and NOISEX-92}. The TIMIT \cite{TIMITgarofolo1993darpa} corpus is selected as another test clean speech, NOISEX-92\cite{NOISEXvarga1993assessment}, and the real-life record noise dataset as the test noise. We use the image source method to generate simulated RIRs as the test RIR set. The room size is set to 5m×4m×3.5m with T60 range is 0.1:0.1:0.5. The locations of the microphone and speaker are randomly in the room with the height range is 1m to 1.5m. We limit the distance of the mic and speaker to 0.2m to 3m. The SNRs are -9dB, -6dB, -3dB, 0dB, 3dB, 6dB, 9dB, 15dB. This test set is much larger and has a wider range of SNR compared to the DNS 2020 test set. The purpose is to test the generalization performance of our model and its performance at low SNRs, and furthermore to determine whether mapping speech or mapping reverse noise is needed in the mask-free approach.

\subsection{Model setup}
All the waveforms are sampled at 16kHz. We use the square root of Hanning window of size 320 with the hop time of 10ms. The optimizer is AdamW. The initial learning rate is set to 0.0034. We used the learning strategy of cosine annealing combined with warmup to reach the maximum of the learning rate in the first 5 epochs. The number of channels of the network is 16. Hyper-parameter $\gamma $ is  0.5. The numbers of blocks in encoder, bottleneck, and decoder are $[d1=1,d2=1,d3=8,d4=4],[m=6],[u1=1,u2=1,u3=1,u4=1]$. By effectively stacking GLFBs, our network is an asymmetric structure. Due to space limitation, we conclude that the experimental result is that the encoder is more important than the decoder, so in the encoder stage, we stack more GLFBs.

\subsection{Evaluation metrics}
Multiple objective metrics are adopted, including narrow-band (NB) and wide-band (WB) perceptual evaluation speech quality (PESQ)  for speech quality, short-time objective intelligibility (STOI) for intelligibility, and SI-SDR  for speech distortion.

\subsection{Ablation study between mask and mask-free}
In this study, we investigated the performance of between mask-based and mask-free methods. The DNS 2020 training set was used, and the synthesized TIMIT test set was used to evaluate the generalization performance of our model under low SNR conditions and unseen speakers. We compared our approach with DCTCRN \cite{DCTCRN}, Cascade DCTCRN, and TaylorSENet. DCTCRN is a masking speech enhancement network that uses STDCT features and won second place in a DNS competition. TaylorSENet is a powerful decoupled-masking model. Additionally, we cascaded the DCTCRN model to enable comparison with our model and multi-stage models. To ensure fairness, all models were trained under the same training configuration. The results are presented in Table \ref{tab2}.

To clarify, the mask method involves connecting a sigmoid function to the network output and then taking the Hadamard product of the noisy STDCT feature with the sigmoid-activated features. Mapping speech involves directly treating the target speech as the learning target in the network output. In contrast, the mapping reverse noise method adds the network output feature to the noisy STDCT feature and then treats the target speech as the learning target. The experimental results indicate that our network achieves better results using the mapping method than the masking method, especially when mapping reverse noise. Furthermore, our network outperforms DCTCRN, Cascade DCTCRN, and TaylorSENet in PESQ, STOI, and SNR metrics.

Once a model has been reasonably trained and has undergone sufficient computation, the masking approach becomes too cautious and fails to fully utilize the model's capabilities. In contrast, the mapping method is less aggressive and appears to be a better fit for this particular model. Interestingly, we observed that in a highly noisy environment, the model performs better by directly learning to reverse the noise.

\subsection{Comparison with the state-of-the-art methods}

We evaluated the proposed SE system on the Interspeech 2020 DNS-Challenge dataset to compare it with other models, and the results are presented in Table \ref{DNS table}. Our MFNet model achieved outstanding performance with a computational volume of only 6.09 GMACs/s. We also conducted an (real time factor) RTF test on the Intel Xeon E5-2680 CPU and the result was 0.236. As few models using the mapping approach were tested on this dataset, we found that the best mapping model is CTSNet, which is a decoupled-based mapping model and an improved version of the TSCN \cite{li2021iDMP} - the winner of the 2021 ICASSP DNS Challenge. CTSNet can be considered a strong competitive model for comparison. To demonstrate the performance of our model, we conducted a horizontal comparison with models from all other methods within a reasonable range of computational complexity for prediction, such as DCCRN, FullSubNet, TaylorSENet and FRCRN. The computational complexity of the FRCRN model is calculated based on our analysis of the website \url{https://modelscope.cn/models/damo/speech_frcrn_ans_cirm_16k/summary} . Our proposed model is highly competitive among these recently proposed models. Our MFNet outperforms the current state-of-the-art mapping network CTSNet by a significant margin. We provide the processed samples, which are available at \url{https://github.com/ioyy900205/MFNet}.

\section{Conclusion}
We present a novel neural network for speech enhancement, called MFNet, which directly learns the real-valued STDCT spectral mapping inspired by the intuitive definition of SA. Our network architecture consists of newly-designed lightweight GLFB modules stacked together to create a simple yet effective single-stage structure capable of modeling global and local information. Using the mapping method, our proposed framework outperforms the current SOTA mapping model on the DNS 2020 test set without reverberation. Overall, our experimental results show that MFNet exhibits superior performance compared to other SOTA models with various alternative approaches. This makes MFNet a promising candidate for practical applications in speech enhancement. In the future, we plan to transform the system into a causal model to facilitate real-world deployment.

\clearpage
\bibliographystyle{IEEEtran}
\bibliography{mybib}

\end{document}